\documentclass[prd,nofootinbib,tightenlines,showkeys,showpacs]{revtex4}
\usepackage{amsfonts}
\usepackage{amssymb}
\usepackage{amsmath}
\usepackage{indentfirst}
\usepackage[latin1]{inputenc}

\begin{document}
\title{Newton's Equation on the kappa space-time and the Kepler problem }

\author{E. Harikumar \footnote{harisp@uohyd.ernet.in} and A. K. Kapoor\footnote{akksp@uohyd.ernet.in}}
\affiliation{School of Physics, University of Hyderabad,\\Central University P O, Hyderabad 500046, India}
\vspace*{1cm}
\begin{abstract}
We study the modification of Newton's second law, upto first order in the deformation parameter $a$, in the $\kappa$-space-time.  We derive the deformed Hamiltonian, expressed in terms of the commutative phase space variables, describing the particle moving in a central potential in the $\kappa$-space-time. Using this, we find the modified equations of motion and show that there is an additional force along the radial direction. Using Pioneer anomaly data, we  set a bond for $a$.  We also analyse the violation of equivalence principle predicted by the modified Newton's equation, valid up to first order in $a$ and use this also to set an upper bound on $a$.

\end{abstract} 


\maketitle

\section{Introduction}
Noncommutative geometry provides a possible way to capture the space-time structure at the Planck scale\cite{connes, sergiod}. Particularly, the quantum gravity effects are expected to drastically affect the symmetries of the space-time itself at the Planck scale\cite{sergiod,sergiod1}. It was also argued that the formation of microscopic black holes would naturally introduce physical cut-off and thus breaks the Lorentz symmetry\cite{thooft}. Noncommutative geometry, treated as a deformation of commutative space-time provides a way of incorporating this cut-off in a natural way. Also, in the limit where the noncommutativity parameter is taken to zero,  Lorentz symmetry is recovered as we would like to have at large length scales. The introduction of fundamental length scale to take into account of the quantum gravity effects breaks the Lorentz symmetry in the usual sense, but can be retained using the Hopf algebra approach\cite{chaichian, wess1,wess1a,majid}.

The Moyal space-time whose coordinates obey $[{\hat X}^\mu, {\hat X}^\nu]=i\theta^{\mu\nu}$, where $\theta^{\mu\nu}$ is a constant and physical models on this space-time have been subject of intense investigations in the last couple of years\cite{rev1, nsew, rev2,pabj,ncgrav,ncgrav1,ncgrav2,ncgrav3}. Noncommutative  space-time with coordinates satisfying a Lie algebra type commutation relation is another class of space-time that is being investigated vigorously.  Fuzzy sphere\cite{madore} which is being studied with various motivations\cite{balbook,balbook1,balbook2,balbook3} is an example for this class.  The $\kappa$-deformed space-time whose coordinates obey
\begin{equation}
 [{\hat x}^i, {\hat x}^j]=0, [{\hat x}^0,{\hat x}^i]=a x^i , (a=\frac{1}{\kappa})\label{kappaco}
\end{equation} 
is another example for Lie algebraic type noncommutative space-time\cite{lukierski,lukierski1,lukierski2,lukierski3}.  $\kappa$-Minkowski space-time naturally appears as the space-time associated with the low energy limit of certain quantum gravity models and the corresponding symmetry transformations known to be related to doubly special relativity (DSR) \cite{dsr} (a modified relativity principle having a fundamental parameter of length dimension  in addition to the velocity of light). This led to the study of $\kappa$-space-time and physics on $\kappa$-space-time in recent times\cite{wess,wessa,wessb,wessb1,glikman,glikman1,glikman2,glikman3,mdljlm,mdljlm1,sm,sma,smb,smc,smd,sme,us,usa,usb,usc,usd}, bringing out its many interesting aspects.

Different field theory models have been constructed on $\kappa$-space-time recently and various aspects of these models have been analysed\cite{lukierski,lukierski1,lukierski2,lukierski3,wess,wessa,wessb,wessb1,glikman,glikman1,glikman2,glikman3,mdljlm,mdljlm1,sm,sma,smb,smc,smd,sme}. It is known that there are many different proposals for Klein-Gordon equation in $\kappa$-space-time \cite{lukierski,lukierski1,lukierski2,lukierski3, wess,wessa,wessb,wessb1,sm,sma,smb,smc,smd,sme}, all satisfying the criterion of invariance under $\kappa$-Poincare algebra.

In last couple of years, the implications of noncommutativity, and in particular that of the Hopf algebra structure of symmetries of such space-time are being studied\cite{jabbari, falomir, vpn, bals,bals1,bals2}. Similar studies to unravel the low energy effects of $\kappa$-deformation have been taken up in\cite{pab,ehms,eh,smksg}.  The effect of  noncommutativity on the classical phase space structure and the resulting changes in classical mechanical systems have been studied in\cite{vergara,vergara1,bmirza,walczyk,fig}.

The noncommutativity of the coordinates deform the symplectic structure of the phase space. This have consequences on the dynamics on such noncommutative space-time. Various examples illuminating this effect have been analysed and bounds on noncommutative parameter have been  obtained \cite{vergara,vergara1,bmirza,walczyk,fig}.

In \cite{vergara,vergara1}, starting with the symplectic structure which is consistent with the Moyal space-time defined by
\begin{equation}
\{{\hat x}^i,{\hat x}^j\}=\theta^{ij},~~\{{\hat x}^i, {\hat p}_j\}=\delta_{j}^i,~~
\{{\hat p}_i,{\hat p}_j\}=0,\label{ncpbs}
\end{equation}
Hamilton's equations were set up.  Note in the above we have used `hat' to emphasise that the coordinates of phase space are those corresponding to the noncommutative space-time. Further, it was assumed in \cite{vergara,vergara1} that the Hamiltonian is of the form
\begin{equation}
H=\frac{{\hat p}\cdot {\hat p}}{2m} +v({\hat x})\label{ham}.
\end{equation}
Using Eqns. (\ref{ncpbs})and (\ref{ham}), the Hamilton's equation of motion was obtained as
\begin{equation}
m{\ddot {\hat x}}^i=-\frac{{\hat x}^i}{{\hat r}}\frac{\partial V}{\partial {\hat r}} +m\epsilon^{ijk}{\dot{\hat x}}_j\Omega_k +m \epsilon^{ijk} {\hat x}_j{\dot\Omega}_k\label{moyalne}
\end{equation}
where $\Omega_i=\frac{1}{{\hat r}}\frac{\partial V}{\partial {\hat r}} \theta_i$ and $\theta^{ij}=\epsilon^{ijk}\theta_k$.

The last two terms in the above equation are the noncommutative corrections to the Newton's equation. These terms are analogue to the inertial force and Coriolis force terms produced 
by non-uniform rotations\cite{vergara,vergara1}. It is clear that these terms break the rotational invariance which can be seen explicitly from the fact that $\{L^i, H\}\ne 0$ where the angular momentum $L^i=m\epsilon^{ijk} {\hat x}_j {\dot{\hat x}}_k$.  But a modified angular momentum $L_\theta$ was shown to be conserved and also generates the rotations in the Moyal space. But in the limit $\theta\to 0$, $L_\theta\to 0$. It was shown that there exist another conserved quantity which in the commutative limit reduces to the familiar angular momentum. 

The effect of the non-inertial terms appearing in Eqn.(\ref{moyalne}) for the orbital motion of planets around the sun have been analysed and it was shown that the noncommutativity leads to the shift of perihelion. Using the observational data of perihelion of Mercury, a bound on the noncommutative parameter was obtained.

In \cite{bmirza}, starting with assumption of the Hamiltonian in the Moyal space as that in Eqn.(\ref{ham}), equation of motion was obtained after mapping the coordinates, momenta and Hamiltonian to corresponding  ones in the commutative space. Here again, rotational invariance was lost and in the approximation where this violation is negligible, shift in the perihelion of Mercury was calculated and used to set a bound on the noncommutative parameter.

In \cite{walczyk}, classical dynamics in noncommutative space were analysed by incorporating the effects of noncommutativity in the deformation of the symplectic structure as in \cite{vergara,vergara1}, but for various types of noncommutative space-times.

In this paper, we study the modification to Newton's second law due to the kappa-deformation, valid upto first order in the deformation parameter, $a$ and its consequences on planetary motion. Here, we start with the deformed energy-momentum relation,  invariant under the $\kappa$-Poincare transformations and derive the Hamiltonian describing the dynamical evolution in the non-relativistic limit. Here, we keep terms up to first order in the deformation parameter. Using this, we derive the Hamilton's equation of motion and analyse problem of a particle moving in the central potential. Unlike the Moyal space-time, here we find that there is no shift in the perihelion of planets in $\kappa$-space-time. But there is a strikingly new effect in the $\kappa$-space-time, an {\it additional} force along the radial direction. This along with the data of Pioneer anomaly set a bound on the deformation parameter and also fix the {\it sign} of the deformation parameter uniquely.  We also show the violation of equivalence principle in the non-relativistic limit of the $\kappa$-space-time and using this also, sets a bound on the deformation parameter.

In the next section, we briefly summarise the $\kappa$-space-time, its symmetry algebra and the deformed energy-momentum relation in this space-time. In section III, we present our main results, viz: derivation of the non-relativistic, $kappa$-deformed Hamiltonian, deformed Hamilton's equation for a particle moving under the influence of $\frac{1}{r}$ potential and effect of $\kappa$-deformation on orbital parameters. We also discuss the violation of equivalence principle here. Our concluding remarks are given in section IV.

\section{$\kappa$-space-time, symmetries and dispersion relation}

In this section we briefly summarise the results of \cite{sm,sma,smb,smc,smd,sme} relevant for the present study.  In these papers, a class of realisations of the coordinates of the $\kappa$-space-time in terms of the commutative coordinates and their derivatives were obtained. Similar results have been obtained in \cite{lukierski,lukierski1,lukierski2, lukierski3,wess,wessa,wessb,wessb1,glikman,glikman1,glikman2,glikman3, mdljlm,mdljlm1} also where Lie algebraic type space-times and field theories on such space-times have been analysed. 
 
We seek for the realisation of the coordinates of the $\kappa$-space-time obeying Eqn.(\ref{kappaco})in terms of the commutative coordinates and their derivatives, i.e.,
\begin{equation}
{\hat x}_\mu=x^\alpha\Phi_{\alpha\mu}(\partial).
\end{equation}
Here 
\begin{equation}
[x^\alpha, x^\beta]=0,~[\partial^\alpha, x^\beta]=\eta^{\alpha\beta}
\end{equation}
where $\eta_{\alpha\beta}=diag(-1,1,1,1)$. 
This realisation defines a unique mapping between the functions of noncommutative space-time to the functions on commutative space-time.  That is, defining the constant function annihilated by $\partial$ as the vacuum, we get
\begin{equation}
F({\hat x}_\varphi)|0>= F(x).\label{map}
\end{equation}
where ${\hat x}_\varphi$ is a specific realisation of ${\hat x}_\mu$, parameterised by $\varphi$(see discussion after eqn.(\ref{cond})).

Subject to the conditions
\begin{eqnarray}
\left[\partial_i, {\hat x}_j\right]=\delta_{ij}\varphi(A),\\
\left[\partial_i, {\hat x}_0\right]=ia\partial_i\gamma(A),\\
\left[\partial_0, {\hat x}_i\right]=0, \left[\partial_0, x_0\right]=\eta_{00},
\end{eqnarray}
where $A=-ia\partial_0$, we find,
\begin{eqnarray}
{\hat x}_i=x_i\varphi(A),\label{real1}\\
{\hat x}_0=x_0\psi(A)+iax_i \partial_i\gamma(A).\label{real2}
\end{eqnarray}

Using Eqns.(\ref{real1}, \ref{real2}) in Eqn.(\ref{kappaco}), we get
\begin{equation}
\frac{\varphi^\prime}{\varphi}\psi=\gamma(A)-1\label{cond}
\end{equation}
where $\varphi^\prime=\frac{d\varphi}{dA}$ satisfying the boundary conditions $\varphi(0)=1, \psi(0)=1, \gamma(0)=\varphi^\prime (0)+1$ and is finite and $\varphi, \psi, \gamma$ are positive functions. The relaisation of ${\hat x}_\mu$ for a specific choice of $\psi$ will be characterised by $\varphi$ and we denote this as ${\hat x}_\varphi$, as used in Eqn.(\ref{map}) above.

Imposing further conditions that the commutators of the Lorentz  generators with the coordinates  ${\hat x}_\mu$ of $\kappa$-deformed space-time are linear in  ${\hat x}_\mu$ and the generators themselves, and these commutators have smooth commutative limit,  lead to just two class of realisations. These are parametrised by $\psi=1$ and $\psi=1+2A$. We consider only the former realisation in the present work (it is this relaisation with $\varphi=e^{-A}$, related to the $\kappa$-Poincare algebra in bi-crossproduct sapce and thus to DSR).

For $\psi=1$, we find,
\begin{eqnarray}
&M_{ij}=x_i\partial_j-x_j\partial_i,&\\
&M_{i0}=x_i\partial_0 \varphi\frac{e^{2A}-1}{2A}-x_0\partial_i\varphi^{-1}+\frac{ia}{2}x_i\nabla^2\varphi^{-2}-ia x_k\partial_k\partial_i\gamma\varphi^{-1}.&
\end{eqnarray}
It is easy to verify that the $M_{\mu\nu}$ defined above satisfy Lorentz algebra but $\partial_0,\partial_i$ do not transform as the components of a 4-vector. It is known that the modified derivative operators (defined using the commutative coordinates and their derivatives)
\begin{eqnarray}
D_i&=&\partial_i e^{-A}\varphi^{-1},\nonumber\\
D_0&=&\partial_0 A^{-1} \sinh A+\frac{ia}{2}\nabla^2 e^{-A} \varphi^{-2},\label{der}
\end{eqnarray}
transform like the components of a four-vector (i.e., $\left[ M_{\mu\nu}, D_\lambda\right]=\eta_{\mu\lambda}D_\nu-\eta_{\nu\lambda}D_\mu$) and satisfy
\begin{eqnarray}
\left[D_\mu, D_\nu\right]=0, \left[M_{\mu\nu}, \Box\right]=0,\\
\left[M_{\mu\nu}, M_{\lambda\rho}\right]=\eta_{\mu\rho}M_{\nu\lambda}+\eta_{\nu\lambda}M_{\mu\rho}-\eta_{\nu\rho}M_{\mu\lambda}-\eta_{\mu\lambda}M_{\nu\rho},
\end{eqnarray}
where
\begin{equation}
\Box=\nabla^2 e^{-A}\varphi^{-2}+2\partial_{0}^2 (1-\cosh A)A^{-2}.
\end{equation}
Note that in the above, 
\begin{equation}
M_{\mu\nu}=({\hat x}_\mu D_\nu-{\hat x}_\nu D_\mu)Z,\label{klorentzgen}
\end{equation}
where $Z=iaD_0+\sqrt{1+a^2D_\alpha D^\alpha}$. Using Eqns.(\ref{real1},\ref{real2}, \ref{der}), $M_{\mu\nu}$ can be expressed in terms of the commutative coordinates and their derivatives.

The $D_\mu$ and $M_{\mu\nu}$ thus define undeformed $\kappa$-Poincare algebra and the corresponding  mass Casimir,  $D_\mu D^\mu=m^2$ in the momentum space is
\begin{equation}
\frac{4}{a^2}\sinh^2(\frac{aE}{2c}) -p_ip_i \frac{e^{-a\frac{E}{c}}}{\varphi^2(a\frac{E}{c})}-m^2c^2 +\frac{a^2}{4}\left[\frac{4}{a^2}\sinh^2(\frac{aE}{2c}) -p_ip_i \frac{e^{-a\frac{E}{c}}}{\varphi^2(a\frac{E}{c})}\right]^2=0,\label{dispersion}
\end{equation}
where $p_i$ are the momentum components corresponding to the commutative coordinates.

\section{ Classical Mechanics on $\kappa$-deformed space}

In this section, we first derive the Hamiltonian describing the time development of a system in the $\kappa$-space-time. This is obtained by taking the non-relativistic limit of the energy-momentum relation given in Eqn.(\ref{dispersion}) and we keep terms of only up to order $a$. Thus the Hamiltonian and hence the equation of motion we get are valid only up to the first order in the deformation parameter $a$.  We also make a specific choice for the function $\varphi(A)$, i.e., we take $\varphi(A)=e^{-A}$. With this choice, the dispersion relation in Eqn.(\ref{dispersion}) becomes that of the $\kappa$-Poincare algebra in the bi-crossproduct basis\cite{majid1,majid1a}, which is known to be related to DSR\footnote{Note that, for $\psi=1$ realisation we are interested in here,  we have $\varphi^\prime(0)=1-\gamma(A)$ which is always a number.  Now, for an arbitrary choice of $\varphi$ (satisfying all the consistency conditions), using Taylor series, we have, up to first order in $a$, $\varphi(A)=1+ia\varphi^\prime(0)\partial_0 $. This differs from the choice $\varphi=e^{-A}=1+ia\partial_0$ only by a numerical factor and hence the general conclusions we arrive at here will be valid, independent of our  choice for $\varphi(A).$}.

From the $\kappa$-deformed dispersion relation in Eqn.(\ref{dispersion}),  we derive, by a straight forward calculation, the Hamiltonian in the non-relativistic limit to be
\begin{equation}
H=\frac{(1-acm)}{2m}~ {\vec p}\cdot {\vec p},\label{kham}
\end{equation}
where we have kept terms up to first order in the deformation parameter $a$. We emphasise here that the momenta $p$ (as well as the coordinates) we use now onwards are the ones in the commutative space-time.

Using Eqns.(\ref{real1}, \ref{real2}) and Eqn.(\ref{map}), with $\varphi(A)=e^{-A}$, we get 
\begin{eqnarray}
{\hat x}_0\to x_0,~~{\hat x}_i\to x_i\varphi(A),\\
{\hat r}^2={\hat x}_i{\hat x}_i\to x_i\varphi(A)x_i\varphi(A)|0>=r^2+\frac{a}{c} r{\dot r},
\end{eqnarray}
where $r=r(t)=\sqrt{x_i x_i}$ and $\dot r=\frac{dr}{dt}$.
Using this, we find
\begin{equation}
\frac{1}{\sqrt{{\hat x}_i{\hat x}_i}}\to\frac{1}{r}-\frac{a\dot r}{2cr^2}
\end{equation}
and thus the potential relevant for the central force problem in the $\kappa$-space-time, valid up to first order in the deformation parameter $a$ is
\begin{equation}
V_{\textrm {eff}}(r)=-\frac{K}{r}-\frac{Ka}{2c}\frac{d~}{dt}\left(\frac{1}{r}\right).\label{kpot}
\end{equation}
We note that the $a$ dependent modification to the potential is a total time derivative.
\subsection{Kepler problem}
From Eqn.(\ref{kham}) and Eqn.(\ref{kpot}), we find the Hamiltonian describing the particle of mass $m$ moving under the influence of the central potential to be (valid up to order $a$)
\begin{eqnarray}
H&=&\frac{(1-acm)}{2m} ~{\vec p}\cdot {\vec p} + V_{\textrm {eff}}(r)\nonumber\\
&=&\frac{(1-acm)}{2m} ~{\vec p}\cdot {\vec p} -\frac{K}{r} +\frac{Ka}{2m c r^3} (x\cdot p),\label{kham1}
\end{eqnarray}
where $K=GMm$ and we have used $p=m{\dot x} + O(a)$. Note here that we have retained only terms up to the order $a$ in the above Hamiltonian. Here we note that the modification to the kinetic term can be captured by a redefinition of the mass as $m_{eff}=\frac{m}{1-amc}$. This should be contrasted with the situation in Moyal case\cite{vergara,vergara1,bmirza} where the kinetic term is not modified at all.

The above Hamiltonian is invariant rotations generated by $L_i=m\epsilon_{ijk}x^j{\dot x}^k$. This is again in contrast to the result in Moyal space-time, where the angular momentum is not a conserved quantity. Here, we also note that replacement of $m$ with $m_{eff}$ in $L_i$ does not affect the rotational invariance of the system described by above Hamiltonian.

The Hamilton's equations of motions are
\begin{eqnarray}
m_{eff}{\dot x}_i&=& p_i + \frac{Ka}{2c}\frac{x_i}{r^3},\\
{\dot p}_i&=&-\frac{Kx_i}{r^3}-\frac{ka}{2mcr^3} x_i+\frac{3Ka}{2m_{eff}} \frac{(x\cdot p)x_i}{r^5},
\end{eqnarray}
which leads to
\begin{equation}
 m_{eff} {\ddot x}_i=-K\frac{x_i}{r^3}=-\frac{\partial V}{\partial x_i}\label{eom}
\end{equation}
where $V=-\frac{K}{r}$. Thus we see here that the Newton's equation gets modified only through the redefinition of the mass. Unlike in the case of Moyal space, there are no additional terms which are analogue to inertial and Coriolis forces.

We re-express the above equation in spherical polar coordinates (rotational invariance allows us to chose $\theta=\frac{\pi}{2}$) as
\begin{eqnarray}
 {\ddot r}- r{\dot \phi}^2=-\frac{K}{m_{eff}}\frac{1}{r^2},\label{reqn}\\
\frac{d}{dt}(m_{eff}r^2{\dot \phi})=0.\label{phieqn}
\end{eqnarray}
Thus we see that the conserved angular momentum is $l_{eff}=(1+amc)l=m_{eff}r^2{\dot \phi}$. This now shows why rotations generated by $L_i$ leaves the system invariant (unlike in the Moyal case) as the $\kappa$-dependent modification to angular momentum is just a scaling of the mass. We also note that, in contrast to the Moyal case, these equations shows that there is no shift in the perihelion of the planets due to the $\kappa$-deformation.

Integrating the Eqn.(\ref{reqn}) give
\begin{equation}
\frac{m_{eff}{\dot r}^2}{2} +\frac{l_{eff}^2}{2m_{eff} r^2} +V=E.\label{energy}
\end{equation}
Note that the conserved energy gets a $a$-dependent modification, which can be related to the commutative expression by redefinitions of $m$ and $l$ to $m_{eff}$ and $l_{eff}$ respectively. Up to first order in $a$, we get
\begin{equation}
 E=E_0(1+amc(1-\frac{V}{E_0})).
\end{equation}
As expected, this will modify the turning points of the orbit and the modified expressions are
\begin{equation}
 r_{\pm}=-\frac{K}{E}\pm\sqrt{\frac{k^2}{E^2}+\frac{2l_{eff}^2}{m_{eff}E}}.
\end{equation}
Using this, we get the semi-major axis of the orbit to be
\begin{equation}
 A=-\frac{K}{E}=A_0(1-\frac{amc}{2}).
\end{equation}
In obtaining the last equality, we have used the approximation $\frac{V}{E}\approx\frac{V}{E_0}=0.5.$ Thus the modified period and angular frequency are given, up to first order in the deformation parameter $a$ as
\begin{equation}
 \tau=\tau_0(1-\frac{amc}{4}),~~\omega=\omega_0(1+\frac{amc}{4}).
\end{equation}
Thus the $\kappa$-deformation do affect the orbital parameters, but there is no shift in the perihelion of planets.

Let us now turn our attention to radial equation. Re-writing Eqn.(\ref{reqn}) by explicitly writing the expression of $m_{eff}$, we find
\begin{equation}
{\ddot r}-r{\dot \phi}^2=-\frac{K}{m}(1-acm)\frac{1}{r^2}\label{reqn1}
\end{equation}
This shows a change in the magnitude of the radial force. This modification depends on the mass of the particle and its distance from sun in addition to $Kac$.

This shows the particle moving in the central potential will experience an acceleration different from that given by the Newton's law of gravitation and its magnitude and direction will depend on the deformation parameter $a$ as well as on $r$. The Pioneer anomaly clearly shows a {\it constant} acceleration of $8.5\times 10^{-10}m/s^2$ directed towards the sun\cite{pioneer,pioneer1}, unaccounted by the Newton's law.  For $|a|=10^{-55}$m(see discussion after Eqn.(\ref{vep})), we find that the force experinaced by the test body of mass $10^3$kg, when it is near Uranus is $1.21 GMc\times 10^{-70}$N and when it is near Saturn is $5GMc\times 10^{-69}$N. Since this change in force is very small, we approximate the additional force due to $\kappa$-deformation to be a constant. Noting that we were working in the units where $\hbar=1$, we see that this additional acceleration experienced by the satellite of mass $10^3$kg when it is near Uranus will set the bound on $|a|\le 10^{-49}m$ while when it is near Mercury, this bound will be $|a|\le 10^{-53}m$. This shows a change in the anomalus acceleration by a factor of $10^{-4}$ as one goes from Uranus to Mercury. We note that to have the acceleration pointed towards sun, $a$ should be a negative quantity. This shows that the right hand side of the second commutation relation given in Eqn.(\ref{kappaco}), should be negative, i.e., $[{\hat x}^0,{\hat x}^i]=-a x^i$(since one would prefer the length parameter $a$ to be positive. Note that this change does not alter any analysis done here). Certainly, this bound is much higher than the Planck length where the $\kappa$-deformation is expected to play a role. Analysing the perihelion shift of Mercury in the Moyal case has set the bound for the corresponding noncommutative parameter as $\theta\le10^{-58}m^2$\cite{vergara,vergara1,bmirza}.

\subsection{ Violation of Equivalence Principle}
It is known that the deformed dispersion relations will lead to the violation of equivalence principle\cite{revep}. We have seen that the modification to the Newton's equation we have obtained in the previous section can be thought of as the corresponding equation in the commutative space with a modified mass. This modification in inertial mass will lead to violation of equivalence principle.  From Eqn.(\ref{eom}) and Newton's law of gravitation, we obtain
\begin{equation}
 \mu_{eff}{\ddot x}=-G\frac{m_g M_g}{r^2},
\end{equation}
where $\mu_{eff}= m M(m+M)^{-1}(1+ac m M(m+M)^{-1})\approx m(1+amc)$.
Thus we find 
\begin{equation}
 {\ddot x}= -G\frac{M_g}{r^2} \frac{m_g}{m} (1-amc) 
\end{equation}
showing that the deviation in the ratio of gravitational to inertial mass is given by
\begin{equation}
 \delta(\frac{m_g}{m})=-\frac{amc}{\hbar}
\end{equation}
where in the last step we introduced $\hbar$ to account for the fact that the calculations were done in the unit where $\hbar=1$.

The experimental limits on violation of equivalence principle\cite{ep,ep1} is one part in $10^{13}$ and this sets
\begin{equation}
 -\frac{amc}{\hbar}\le 10^{-13}\label{vep}
\end{equation}
implying $a$ is negative and $|a|\le 10^{-55}m$ for a body mass $1 Kg$. This bound is $10^{-6}$ times of that obtained from Pioneer anomaly(where we have taken $|a|\le 10^{-49}$m, obtained when the test body is near Uranus). Note that in both cases, $a$ has to be negative.
\section{Conclusion}

We have studied the modification to Newton law of motion due to $\kappa$-deformation of the underlying space-time. Starting from the dispersion relation invariant under $\kappa$-Poincare transformation, we derived the (non-relativistic) Hamiltonian, valid up to first order in the deformation parameter. We have also showed that the modification to $\frac{1}{r}$ potential due to $\kappa$ deformation is can be written as a total time derivative. The Hamiltonian describing a particle in a central potential do have rotational invariance (which is lost in the case of Moyal noncommutative space-time).  This rotational invariance is not unexpected since the space coordinates commute among themselves in $\kappa$-space-time. We also note the $\kappa$-deformation modify the boost sector while leave the rotation sector unchanged\cite{majid1,majid1a}. We also notice that the rotation generators defined in Eqn.(\ref{klorentzgen}), with $\varphi=e^{-A}$ is exactly same as that in the commutative space-time. Thus the algebra of these generators are also not modified. This explains why the Hamiltonian in Eqn.(\ref{kham1}) has rotational invariance. Using these results, we have obtained the equation of motion for a particle moving in the central potential. We have shown that the orbital parameters do get $a$ dependent changes while there is no shift in the perihelion. This is in contrast to the results obtained in the Moyal case\cite{vergara,vergara1,bmirza}. There are no inertial and/or Coriolis terms in the equation of motion unlike in the case of Moyal space.

We see that the $\kappa$-deformation modified both kinetic and potential terms of the Hamiltonian in Eqn.(\ref{kham1}). While the change in kinetic energy term is just a scaling of the coefficient, potential term is modified by the addition of a new term. But, up to first order in $a$, this additional term is a total time derivative\footnote{The second order correction to potential is not a total derivative.}.

The modified kinetic term in the Hamiltonian lead to the change in the equations of motion whereas that in the potential (up to first order in $a$), do not contribute to any new terms in the equation of motion.

The Eqn.(\ref{reqn}) shows an additional force along the radial direction. Pioneer anomaly  indicates existence of such a force in the solar system. Using the Pioneer data, we find that the sign of the right hand side of the non-vanishing commutator in Eqn.(\ref{kappaco}) is negative and the deformation parameter is bounded by an upper value of $10^{-49}m$.  We have also analysed the violation of equivalence of principle by evaluating the deviation of the ratio between inertial and gravitational masses. This also sets the sign of the commuatator in Eqn.(\ref{kappaco}) to be negative and impose a bound, $|a|\le10^{-52}m$ for a body of mass $1gm$.

\end{document}